\def\ugr{\lower4pt \hbox{$\buildrel > \over \sim$}}
\def\ukl{\lower4pt \hbox{$\buildrel < \over \sim$}}
\documentstyle[epsf, rotate]{aipproc}

\begin{document}
\title{Spectral modelling of gamma-ray blazars}

\author{M. B\"ottcher, H. Mause and R. Schlickeiser}
\address{Max-Planck-Institut f\"ur Radioastronomie, Postfach 20 24, 
D -- 53 010 Bonn, Germany}

\maketitle

\begin{abstract}
We present model calculations reproducing broadband 
spectra of $\gamma$-ray blazars by a relativistic leptonic 
jet, combining the EIC and the SSC model. To this end, the 
evolution of the particle distribution functions inside a
relativistic pair jet and of the resulting photon spectra 
is investigated. Inverse-Compton scattering of both
external (EIC) as well as synchrotron photons (SSC)
is treated using the full Klein-Nishina cross-section
and the full angle-dependence of the external photon
source. We present model fits to the broadband 
spectra of Mrk~421 and 3C279 and the X-ray and $\gamma$-ray
spectrum of PKS~1622-297. We find that the most plausible way 
to explain both the quiescent and the flaring states of these 
objects consists of a model where EIC and SSC dominate the 
observed spectrum in different frequency bands. For both 
Mrk~421 and 3C279 the flaring states can be reproduced by a 
harder spectrum of the injected pairs. 
\end{abstract}

\section*{Introduction}

The discovery of many blazar-type AGNs (e. g., Hartman 
et al. 1992) as sources of high-energy $\gamma$-ray 
radiation, dominating the apparent luminosity, has 
revealed that the formation of relativistic jets and 
the acceleration of energetic charged particles, that 
generate nonthermal radiation, are key processes to 
understand the energy conversion process. Emission 
from relativistically moving sources is required to 
overcome $\gamma$-ray transparency problems implied 
by the measured large luminosities and short time 
variabilities (for review see Dermer \& Gehrels 1995).

Dermer, Miller \& Li (1996) have recently inspected the 
acceleration of energetic electrons and protons in the 
central AGN plasma by comparing the time scales for 
stochastic acceleration with the relevant energy 
loss time scales. Their results demonstrate that with
reasonable central AGN plasma parameter values low-frequency 
turbulence can energize protons to TeV and PeV energies 
where photo-pair and photo-pion production are effective 
in halting the acceleration (Sikora et al. 1987, 
Mannheim \& Biermann 1992). Once the accelerated 
protons reach the thresholds for the latter processes 
they will generate plenty of secondary electrons and 
positrons of ultrahigh energy which are now injected 
at high energies into the acceleration scheme. The 
further fate of the secondary particles depends 
strongly on whether they find themselves in a compact 
environment set up by the external accretion disk, 
or not. Secondary particles within the photosphere 
will initiate a rapid electromagnetic cascade which has 
been studied by e.g. Mastichiadis \& Kirk (1995), which 
might even lead to runaway pair production and associated 
strong X-ray flares (Kirk \& Mastichiadis 1992), and/or 
due to the violent effect of a pair catastrophy (Henri, 
Pelletier \& Roland 1993) ultimately lead to an explosive event 
and the emergence of a relativistically moving component filled 
with energetic electron-positron pairs.

It is the purpose of the present investigation to follow the 
time evolution of the relativistic electrons and positrons 
as the emerging relativistic blob moves out. We generalize 
the approach used in earlier work (Dermer \& Schlickeiser 
1993, Dermer, Sturner \& Schlickeiser 1996) by accounting for 
Klein-Nishina effects and including external inverse-Compton 
scattering as well as synchrotron self-Compton scattering 
self-consistently.

\section*{Simulations of the temporal evolution of particle and
photon spectra}

We assume the relativistic pairs to be isotropically distributed 
in a spherical volume of radius $R_B$, located at height $z_i$
above the accretion disk plane. The minimum Lorentz 
factor of the pairs at the time of injection is expected 
to be in the range of the threshold value of the 
primary protons' Lorentz factor for photo-pair production.
We find this threshold at $\gamma_{1\pm} \sim 10^3$. 
The pair distribution above this cutoff basically reflects 
the acceleration spectrum of the protons which can extend 
up to $\gamma_{2\pm} \sim 10^6$. The lack of a significant
cut-off at high photon energies due to $\gamma$-$\gamma$
pair production in the EGRET spectra of many $\gamma$-ray
blazars and in particular the detection of TeV 
$\gamma$-rays from Mrk~421 suggests that a new jet component
must be produced/accelerated outisde the $\gamma$-ray photosphere 
for TeV photons. We find this photosphere due 
to the interaction of $\gamma$-rays with accretion disk 
radiation to be located around $z \approx 5 \cdot 10^{-3} \,
M_8^{-1}$~pc where $M_8$ is the mass of the central black 
hole in units of $10^8 \, M_{\odot}$. 

The blob is assumed to move outward perpendicularly 
to the accretion disk plane with velocity $c \, \beta_{\Gamma} 
= c \, \sqrt{1 - 1/\Gamma^2}$. The cooling mechanisms we take 
into account are inverse-Compton scattering of accretion disk 
photons (EIC), synchrotron and synchrotron-self-Compton (SSC) 
losses to arbitrarily high order. For comparison
to observations, we calculate the time-averaged 
emission, since the integration times of present-day
$\gamma$-ray observing instruments are much longer than the 
cooling time-scales resulting from our simulations.

\begin{figure}
\epsfysize=4cm
\hskip 2cm
\rotate[r]{
\epsffile[90 50 600 300]{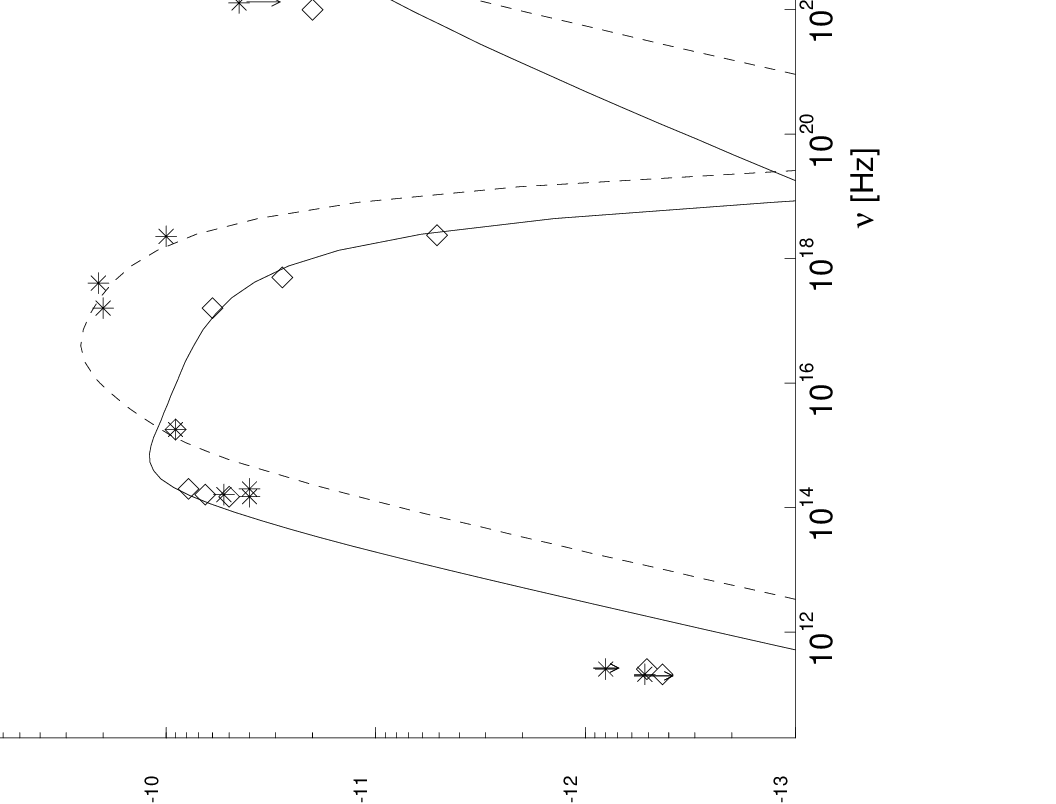}}
\caption{Fit to the broadband spectrum of Mrk 421 in its quiescent state 
(squares; solid line) and during the May 1994 flare (stars; dashed line)}
\end{figure}

\section*{Model calculations}

We used our code to fit the spectra of three $\gamma$-ray blazars,
namely the flat-spectrum radio quasar 3C279, the BL-Lac object 
Mrk~421 and the quasar PKS~1622-297. Throughout our calculations, 
we assumed $H_0 = 75$~km~s$^{-1}$~Mpc$^{-1}$, $q_0 = 0.5$, 
and $\Lambda = 0$. The fits to Mrk~421, as shown in Fig. 1,
are discussed in detail in B\"ottcher et al. (1997).

\subsection*{The quasar 3C279}

The quasar 3C279 was observed in a broad-band campaign by Hartman
et al. (1996) during an outburst in 1991 June and in a more
quiescent state in 1991 September -- October. For the latter
phase, unfortunately, no simultaneous observations in the
infrared -- optical -- UV band were available. The typical
flare timescale of this object is several days, which restricts
the size of the emitting region to $R_B \ukl D \cdot 10^{17}$~cm
where $D = (\Gamma \, [1 - \beta_{\Gamma}\cos\theta_{obs}])^{-1}$
is the Doppler factor. Our fits to the quiescent and the flaring 
phase of 3C279 are shown in Fig. 2. The parameters for the 
quiescent state were

$$
\vbox{
\halign to 0.5\hsize{ \hfil $#$ \tabskip = 0pt & = $#$ \tabskip = 1cm plus 1cm
                      \hfil & \hfil $#$ \tabskip = 0pt & = $#$ \hfil \cr
\gamma_{1\pm} & 200                 & \gamma_{2\pm} & 7 \cdot 10^4 \cr
s & 2.5                             & n & 3 \, {\rm cm}^{-3} \cr
R_B & 3 \cdot 10^{17} \, {\rm cm}   & B & 0.1 \, {\rm G} \cr
L & 10^{46} \, {\rm erg \, s}^{-1}  & M & 5 \cdot 10^8 \, M_{\odot} \cr
z_i & 6 \cdot 10^{-2} \, {\rm pc} & \Gamma & 15 \cr
\theta_{obs} & 2^0 & \omit & \omit \cr } } $$
where $\gamma_{1,2 \pm}$ are the cutoffs of the initial pair
distributions, $s$ is the initial spectral index of the particle 
spectra ($n(\gamma) \propto \gamma^{-s}$), $z_i$ is the height
of the injection site above the accretion disk plane, $L$ is the
total luminosity of the accretion disk and $M$ is the masss of the
central black hole.

\begin{figure}
\epsfysize=4cm
\hskip 2cm
\rotate[r]{
\epsffile[90 50 600 300]{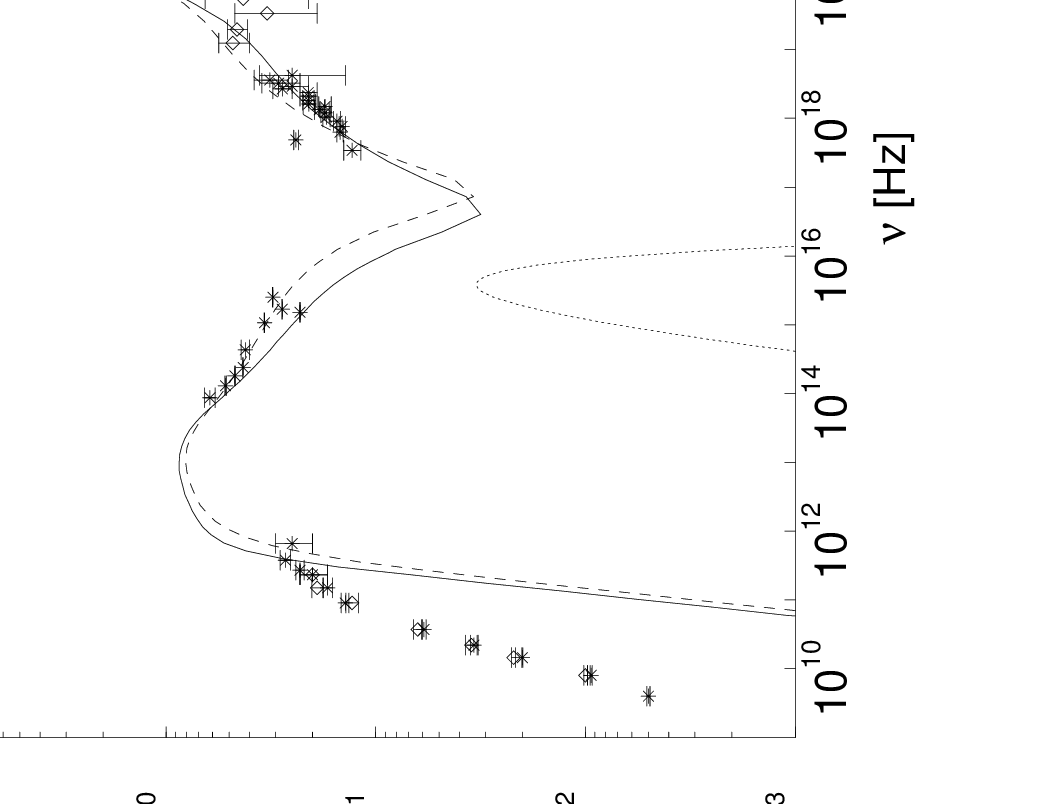}}
\caption{Fit to the broadband spectrum of 3C279 in its quiescent state (squares;
solid line) during 1991 September -- October, and during the flare of
1991 June (stars; dashed line). Dotted curve: accretion disk spectrum}
\end{figure}

The flaring state could be reproduced by a harder injection spectrum
with $s = 2.4$, $\gamma_{1\pm} = 300$ and $\gamma_{2\pm} = 8 \cdot 10^4$.

\subsection*{The quasar PKS 1622-297}

Recently, Mattox et al. (1997) have reported an intense $\gamma$-ray
flare of the quasar PKS~1622-297 during which the X-ray and the
$\gamma$-ray spectrum can be represented by two different 
power-laws. Our model calculation, which is illustrated in 
Fig. 3, demonstrates that the combined SSC/EIC model is well 
suited to reproduce such a two-component spectrum. 

The parameters chosen for the fit shown in Fig. 3 are:
$$
\vbox{
\halign to 0.5\hsize{ \hfil $#$ \tabskip = 0pt & = $#$ \tabskip = 1cm plus 1cm
                      \hfil & \hfil $#$ \tabskip = 0pt & = $#$ \hfil \cr
\gamma_{1\pm} & 400                 & \gamma_{2\pm} & 7 \cdot 10^4 \cr
s & 2.7                             & n & 420 \, {\rm cm}^{-3} \cr
R_B & 10^{17} \, {\rm cm}           & B & 0.08 \, {\rm G} \cr
L & 10^{46} \, {\rm erg \, s}^{-1}  & M & 10^8 \, M_{\odot} \cr
z_i & 2.5 \cdot 10^{-2} \, {\rm pc} & \Gamma & 15 \cr
\theta_{obs} & 4^0 & \omit & \omit \cr } } $$

\begin{figure}
\epsfysize=3.5cm
\hskip 2.3cm
\rotate[r]{
\epsffile[90 50 600 300]{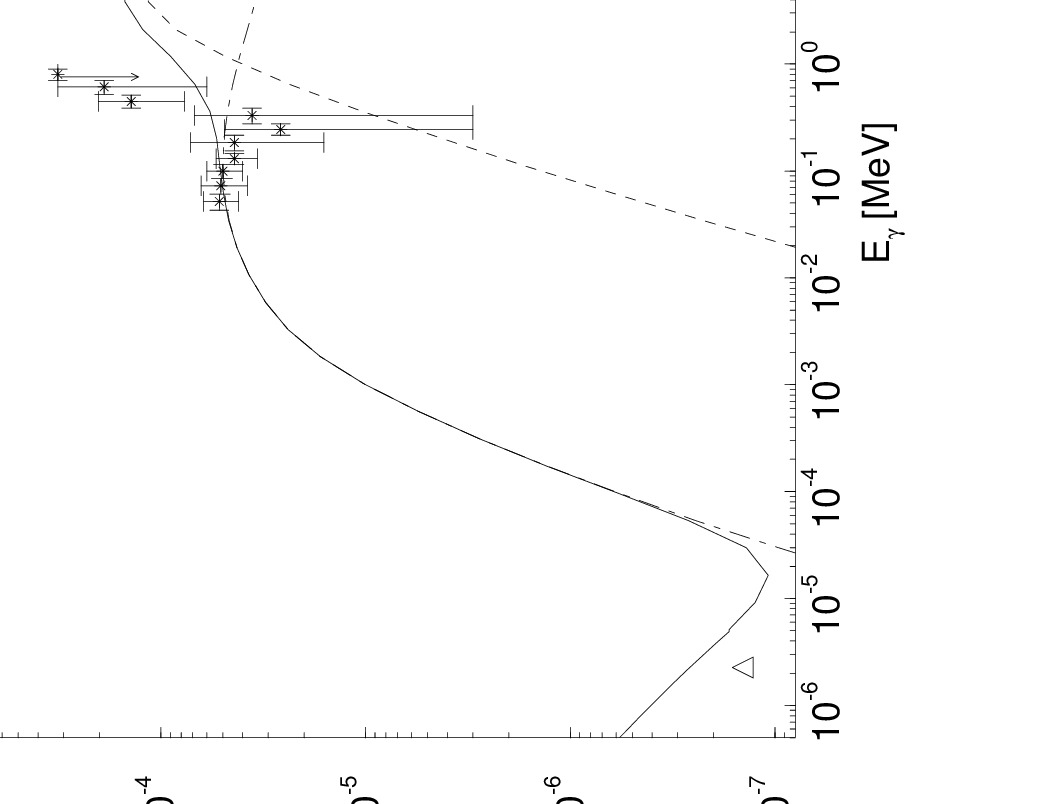}}
\caption{Fit to the X-ray and $\gamma$-ray spectrum of PKS~1622-297
during the flare reported by Mattox et al. (1997); the optical data 
point is from non-simultaneous observations and does not correspond
to the flare state of PKS~1622-297. Dashed: EIC, dot-dashed: SSC}
\end{figure}

\end{document}